# Is the Supernova Remnant RX J1713.7-3946 a Hadronic Cosmic Ray Accelerator?

Yousaf M. Butt[1], Diego F. Torres[2,3], Jorge A. Combi[3], Thomas Dame[1], Gustavo E. Romero[3]

[1]*Harvard-Smithsonian Center for Astrophysics, 60 Garden St., Cambridge, MA, USA.*

[2]*Department of Physics, Princeton University, Princeton, NJ, USA*

[3]*Instituto Argentino de Radioastronomía, C.C. 5, 1894, Villa Elisa, Buenos Aires, ARGENTINA*

**The non-thermal supernova remnant (SNR) RX J1713.7-3946 has recently been shown to be a site of cosmic ray (CR) electron acceleration to TeV energies (Muraishi et al., 2000). Here we present evidence that this remnant is also accelerating cosmic ray *nuclei*. Such energetic nuclei can interact with ambient interstellar gas to produce high energy gamma-rays via the decay of neutral pions. We associate the unidentified *EGRET* GeV gamma-ray source, 3EG J1714-3857, with a very massive (~3×10$^5$ M$_o$) and dense (~500 nucleons cm$^{-3}$) molecular cloud interacting with SNR RX J1713.7-3946. Direct evidence for such interaction is provided by observations of the lowest two rotational transitions of CO molecules in the cloud; as in other clear cases of interaction, the CO(*J*=2→1)/CO(*J*=1→0) ratio is significantly enhanced. Since the cloud is of low radio and X-ray brightness, electrons cannot be responsible for the bulk of the GeV emission there. A picture thus emerges where both electrons and nuclei are being accelerated by the SNR: whereas the relativistic electrons dominate the *local* non-thermal radio, X-ray and TeV emission, the shock accelerated CR protons and ions (hadrons) are exposed through their interactions in the *adjacent* massive cloud, leading to the observed GeV emission via the gamma-decay of neutral pions. Such a scenario had been anticipated by Aharonian, Drury and Völk (1994).**

Subject headings: acceleration of particles --- cosmic rays --- ISM: clouds --- supernova remnants (RX J1713.7-3946, G347.3-0.5)

# 1. INTRODUCTION

The question over the origin of cosmic rays (CR) has persisted ever since they were first detected in 1912 by Victor Hess using a balloon-borne electroscope device (Hess, 1912). Despite the fact that there is now broad consensus that the shocks in the expanding blast waves of supernova remnants (SNRs) accelerate the bulk of the CR ions up to energies of ~300 TeV per nucleon, and possibly to even ~$10^5$ TeV per nucleon (Bell & Lucek, 2001), direct evidence of this scenario has so far eluded observers. The best way to identify such energetic CR accelerators is to look for the associated high-energy gamma-rays produced at the sources; however, since CR electrons and nuclei can both generate gamma-rays at the acceleration sites, it has not yet been possible to unambiguously associate the detected gamma-radiation with sources of CR *nuclei*, specifically. In fact, only very recently have ground-based imaging air Čerenkov gamma-ray telescopes provided direct evidence for the presence of highly relativistic TeV energy particles, most likely electrons, at the shocks of three shell-type SNR's: SN1006 (Tanimori et al., 1998), Cassiopeia A (Aharonian et al., 2001) and RX J1713.7-3946 (also known as G347.3-0.5; Muraishi et al., 2000), the remnant considered here. Even though previous reports (eg. Sturner et al., 1996; Esposito et al., 1996; Gaisser et al., 1998; Aharonian & Atoyan, 1999; Combi et al., 1998, 2001) have suggested that the gamma-ray emission seen in the directions of some SNRs could be due to hadronic interactions, it has not been possible to rule out energetic electrons or the nearby pulsars as the dominant source of the detected radiation (eg. Brazier et al., 1996; De Jager, & Mastichiadis, 1997; Gaisser et al., 1998).

## 2. 3EG J 1714-3857 AND SNR RX J1713.7-3946

Prompted by the close association of the GeV *EGRET* source 3EG J1714-3857 (Hartman et al., 1999) with the TeV gamma-ray emitting SNR RX J1713.7-3946, we have investigated the possibility that this remnant is also accelerating nuclei, *in addition to the known CR electron acceleration taking place there* (Muraishi et al., 2000). As Figure 1 shows, there are two massive and dense molecular clouds lying adjacent to the SNR; one of which, Cloud A, immediately abuts the blast wave region of the SNR (Slane et al., 1999). This cloud is also partially within the inner 50% confidence location contour of 3EG J1714-3857 (Hartman et al., 1999). Both clouds have a mean Local Standard of Rest (LSR) velocity of $-94$ km/sec and an inferred kinematic distance of $6.3 \pm 0.4$ kpc. The mass of Cloud A, is determined from CO observations (Bronfman et al., 1989) to be $(3 \pm 0.3) \times 10^5$ $M_o$ and its mean density ~500 nucleons cm$^{-3}$. Cloud B has a mass of $(2.8 \pm 0.3) \times 10^5$ $M_o$, and mean density of ~660 nucleons cm$^{-3}$.

As the blast wave of RX J1713.7-3946 overtakes Cloud A the shock-accelerated protons and ions collide with the resident nuclei and produce neutral pions which then promptly gamma-decay ($\pi^o \rightarrow \gamma\gamma$), illuminating the cloud at GeV energies; such a scenario has been anticipated by several authors (eg. Montmerle, 1979; Aharonian, Drury & Volk, 1994; Dorfi, 1991, 2000). Indeed, a recent detailed analysis of the broadband *electronic* emissions of RX J1713.7-3946 by Ellison et al. (2001) directly supports our findings by suggesting that 25-50% of the forward shock kinetic energy is likely being taken up in accelerating ions to relativistic energies of up to ~70 TeV/nucleon in this remnant. Strong evidence that the shock front of RX J1713.7-3946 has overtaken, and is interacting with Cloud A, is provided by the enhanced intensity ratio of the two lowest rotational transitions of the CO molecules in the cloud, as already noted by Slane et al. (1999). This ratio, $R=\{CO(J=2\rightarrow1)/CO(J=1\rightarrow0)\}$, is typically ~0.7 in the Galactic plane (Sakamoto et al., 1995), but is known to be enhanced in shocked molecular gas interacting with SNRs (eg. Seta et al., 1998). We examined this ratio, $R$, at all LSR velocities over a $2^{o}\times1^{o}$ region centered roughly on Cloud A, using $CO(J=1\rightarrow0)$ data from Bronfman *et al.* (1989), and unpublished $CO(J=2\rightarrow1)$ data from the University of Tokyo 0.6m telescope at La Silla, Chile, kindly provided by T. Handa and T. Hasegawa. As Figure 2 shows, of all the 781 ratios measured, 2 of the 3 highest (top 0.5% percentile), with $R\sim2.4\pm0.9$, were found on or very close to Cloud A in both position and velocity. In contrast, no enhanced $CO(J=2\rightarrow1)/CO(J=1\rightarrow0)$ ratio was observed in the vicinity of Cloud B.

We have calculated the expected gamma-ray luminosity for the proposed scenario using the following information from Slane et al. (1999): supernova explosion energy = $E_{SN}$ = (1.7-2.2)×10$^{51}$ ergs; distance to the SNR = 6.3 ± 0.4 kpc; unshocked ambient density, $n_o$ = 0.01 − 0.3 cm$^{-3}$; together with the cloud data extracted above. Using the Sedov solution we calculate the age of the SNR to be in the range (2.5 − 13.4)×10$^4$ yrs. The total gamma-ray luminosity is divided between that from the hadronic interactions intrinsic to the SNR, and that due to the enhanced probability of hadronic interactions in the high target density medium of Cloud A:

$$F_{tot}(E>100\text{MeV}) = F_{snr}(E>100\text{MeV}) + F_{cloud\,A}(E>100\text{MeV}) \qquad [1]$$

We may evaluate the first term as (Drury et al., 1994):

$$F_{snr}(E>100\text{MeV}) \sim 4.4 \times 10^{-7}\ \theta\,E_{51}\,D^{-2}_{kpc}\,n_o\ \text{photons cm}^{-2}\text{ sec}^{-1} \qquad [2]$$

where $\theta$ is the fraction of the total supernova energy converted to cosmic ray energy; $E_{51}$ is the supernova explosion energy in units of $10^{51}$ erg; and $D_{kpc}$ is the distance in kpc. If the start of the Sedov phase is taken at ~2400 years, and $\theta \sim 0.5$ (eg. Morfill et al., 1984) then substituting the numerical values for the various quantities yields the intrinsic GeV luminosity of the SNR to be in the range:

$$F_{snr}(E>100\text{MeV}) = (0.1-3)\times 10^{-9} \text{ photons cm}^{-2} \text{ sec}^{-1} \qquad [3]$$

The second term represents the contribution to the GeV flux from the SNR amplified CR bombardment of Cloud A and is given by (Aharonian & Atoyan, 1996):

$$F_{cloud\ A}(E>100\text{MeV}) = 2.2\times 10^{-7} M_5 D^{-2}_{kpc} k_s \text{ photons cm}^{-2} \text{ sec}^{-1} \qquad [4]$$

where $M_5$ is the mass in units of $10^5 M_o$ and $k_s$ is the cosmic ray enhancement factor, ie. the ratio of the CR energy density in the vicinity of the SNR to that measured near the Sun. [We adopt a gamma-ray emissivity, $q(E>100\text{MeV})=2.2\times 10^{-25}$ photons (H-atom)$^{-1}$ sec$^{-1}$, (Dermer, 1986)]. Using Morfill et al. (1984), we find $24 < k_s < 36$ for the SNR at the current epoch. Thus, the GeV luminosity of Cloud A is evaluated to be in the range $(3.1 - 7.6)\times 10^{-7}$ photons cm$^{-2}$ sec$^{-1}$, *and dominates, by more than two orders of magnitude, the GeV flux produced by the SNR itself*. Since the intrinsic SNR contribution can thus be neglected, the total expected hadronically generated gamma-ray luminosity is simply the same as that from Cloud A:

$$F_{tot}(E>100\text{MeV}) \sim F_{cloud\ A}(E>100\text{MeV}) = (3.1 - 7.6) \times 10^{-7} \text{ photons cm}^{-2} \text{ sec}^{-1} \qquad [5]$$

That this predicted flux is fully consistent with the measured value, $(4.36 \pm 0.65)\times 10^{-7}$ photons cm$^{-2}$ s$^{-1}$ (Hartman et al., 1999), further supports the proposed nucleonic source of the detected gamma rays. Also, the fact that the *EGRET* source is coincident with the molecular cloud, and not the SNR itself, is fully in agreement with our calculation of the expected relative fluxes from these two sources. In addition, the spectral index of the GeV source, $\Gamma = -2.3 \pm 0.2$ (Hartman et al., 1999), is in tune with that expected from the hadronic interactions of a *source* CR population (Fields et al., 2001).

To be certain of a nucleonic source of the detected GeV flux from Cloud A, however, it is crucial to eliminate the alternative, electromagnetic origin of the gamma-rays. We show that the *non*-detection of Cloud A in the radio band (Slane et al., 1999; Ellison et al., 2001) rules out the possibility that electrons are contributing significantly to the GeV luminosity *of the cloud*. At the high particle densities of Cloud A, the contribution of the electron IC process to the GeV luminosity can be neglected in comparison to the electron bremsstrahlung process (eg. De Jager & Mastichiadis, 1997). However, the electron flux needed to explain the intensity of the measured GeV emission via electron bremsstrahlung in the cloud material would produce an enhanced radio emission by the synchrotron mechanism which far exceeds the measured values. The expected ratio of gamma-ray (>100 MeV) electron bremsstrahlung flux to the radio synchrotron flux may be expressed:

$$\frac{F(E > 100\,\mathrm{MeV})}{F(v)_{\mathrm{Jy}}} = \frac{4.3 \times 10^{-21}}{c(p)} n_{\mathrm{cm}^{-3}} B_{\mu G}^{-(1+p)/2} v_{\mathrm{Hz}}^{(p-1)/2} \quad \mathrm{Jy}^{-1}\,\mathrm{cm}^{-2}\,\mathrm{s}^{-1} \;, \quad [6]$$

where,

$$c(p) = 10^{-5(1+p)} (3.2 \times 10^{15})^{(p-1)/2} (p-1)^2 a(p),$$

*a(p)* is given in Longair (1994), and *p*=2.3 is the spectral index of the electron population, $N_e(E) \sim E^{-2.3}$.

Then, using the physical parameters of Cloud A, together with an assumed magnetic field of ~25 $\mu$G [a conservative estimate for molecular clouds (Crutcher et al., 1987)], we can calculate that the predicted radio luminosity at 843 MHz – *under the assumption of an electronic origin of the GeV flux* – would be ~40 Jy. Since this flux is about an order of magnitude larger than the *upper limit* derived from the non-detection of Cloud A at this frequency (Slane et al., 1999), we conclude that no significant part of Cloud A's GeV radiation could be due to electronic processes unless the magnetic field strength in the molecular cloud were as low as the ISM value of ~5$\mu$G. Furthermore, were the GeV flux of 3EG J1714-3857 of electronic origin, Cloud A would outshine even the radio-brightest NW rim of the remnant, which is found to be emitting at only 4±1 Jy at 1.36 GHz (Ellison et al., 2001). There is, of course, a relativistic bremsstrahlung contribution from *secondary* electrons and positrons produced by the decay of *charged* pions ($\pi^{\pm} \rightarrow \mu^{\pm} \rightarrow e^{\pm}$) which are also generated in the hadronic interactions, but this gamma-ray intensity is expected to be more than an order of magnitude lower than the $\pi^0 \rightarrow \gamma\gamma$ flux above 100 MeV (Berezinskii et al., 1990).

## 3. OTHER SOURCES IN THE FIELD

There are also two other SNRs projected within 3EG J1714-3857's 95% contours, CTB37A&B. However, since these SNRs are more distant (11.3 kpc), and because their *maximum possible* interacting cloud mass (Reynoso & Mangum, 2000) is measured to be an order of magnitude less than Cloud A's, their contribution to the GeV luminosity is less than $2\times10^{-8}$ photons cm$^{-2}$ sec$^{-1}$, or less than 5% of the measured 3EG J1714-3857 flux.

The two pulsars within the 95% confidence location contours of this *EGRET* source, PSR J1715-3903 and J1713-3844 (Manchester et al., 2001), can also be eliminated as the source of the bulk of the measured GeV flux. PSR J1713-3844 at ($l,b$)=(348.10,+0.21) is a long-period pulsar (P=1.60011 sec) whose spin-down luminosity is two orders of magnitude below that needed to account for 3EG J1714-3857. Although spinning faster, PSR J1715-3903 at ($l,b$) = (348.10, -0.32), is still not energetic enough to be responsible for 3EG J1714-3857: it has a period P=0.27848 sec; the dispersion measure indicates a distance of $d$=4.8 kpc and the observed period derivative of $37.688\times10^{-15}$ implies a spin-down luminosity of $\dot{E}\sim7\times10^{34}$ erg s$^{-1}$, for a standard neutron star moment of inertia $I=10^{45}$ g cm$^2$. Thus, $\dot{E}/d^2 \sim 3\times10^{33}$ erg s$^{-1}$ kpc$^{-2}$ which is more than two orders of magnitude below the lowest value among the confirmed gamma-ray pulsars[1] (Kaspi et al., 2000). We thus conclude that the pulsar J1715-39 is not responsible for the bulk of the GeV emission of 3EG J1713.7-3946. This conclusion is supported by the lack of any X-ray counterpart of PSR J1715-39 in the available archival X-ray databases.

Lastly, the *EGRET* source 3EG J1713.7-3946 is not coincident with any other candidate gamma-ray sources such as OB associations, Wolf-Rayet or Of stars (Romero et al., 1999; Torres et al., 2001a). The analyses of both Tompkins (1999) and Torres et al. (2001b) also shows this source to be non-variable, as should be the case for an interacting SNR.

---

[1] We do not consider here PSR B1055-52 since there is an open controversy regarding the distance to this pulsar – see Combi et al. (1997), Romero (1998), Mc Laughlin & Cordes (2000) and Torres et al. (2001c) for discussions.

9## 4. CONCLUSIONS

We have argued that the unidentified *EGRET* source 3EG J1714-3857 (Hartman et al., 1999) results predominantly from the gamma-rays produced by nuclei accelerated by SNR RX J1713.7-3946 interacting with those resident in the dense and massive molecular cloud immediately abutting the remnant. A recent analysis of the *electronic* emissions of this remnant by Ellison et al. (2001) directly supports our proposal by suggesting that 25-50% of the forward shock kinetic energy is likely being taken up in accelerating ions to CR energies.

However, it should be noted that a number of theoretical assumptions are built into the simple models of the literature cited in our analysis of the hadronic gamma-ray production (Morfill et al., 1984; Aharonian et al., 1994; Drury et al., 1994; Aharonian & Atoyan, 1996). For instance, the diffusion and confinement of the protons in the dense and magnetized media of molecular clouds is a complicated problem whose detailed analysis is beyond the scope of this letter (see, eg. Zweibel & Shull, 1982; Berezinskii et al., 1990; Dogiel & Sharov, 1990; Chandran, 2000). The large angular size of the *EGRET* error box [Fig. 1] also leaves open the possibility that some other, as yet unidentified, source could also be contributing significant gamma-ray flux. Observations of this region with the upcoming higher sensitivity and spatial resolution satellite-based GeV telescopes, such as *AGILE* and *GLAST*, will thus be very important. We also propose that Cloud A of RX J1713.7-3946 be a high priority target for *CANGAROO-III* and *HESS*, the forthcoming high-sensitivity ground-based TeV Čerenkov telescope arrays in the southern hemisphere. Such observations of very high energy photons from Cloud A could directly probe the maximum proton energy, $E_{p-max}$, accelerated by RX J1713.7-3946, since the hadronic gamma-ray spectrum begins steepening at $E_\gamma \sim 0.1\, E_{p\text{-}max}$ and is cut-off at $E_\gamma \sim E_{p\text{-}max}$ (eg. Naito & Takahara, 1994).

In conclusion, the facts that TeV energy cosmic ray electrons are accelerated in SNR RX J1713.7-3946 (Muraishi et al., 2000); that the abutting cloud material is inordinately excited; that the cloud region is of low radio and X-ray brightness; that the GeV luminosity is non-variable and in quantitative agreement with that expected from $\pi^o$ gamma-decays; that the spectral index is as expected for an hadronic CR source population; and, lastly, that there are no other known candidate sources within the 95% location contours of 3EG J1714-3857 capable of explaining the GeV flux, all suggest that this *EGRET* source is the gamma-ray signature of accelerated nuclei from SNR RX J1713.7-3946 interacting with those of the neighboring dense and massive molecular cloud.

We thank T. Handa and T. Hasegawa for kindly providing the CO($J$=2→1) data, obtained at the University of Tokyo 0.6m telescope at La Silla, Chile. We are grateful to Patrick Slane for alerting us to RX J1713.7-3946 and to Don Ellison, Sera Markoff and David Thompson for useful discussions and information. The *EGRET* and HEASARC archives at the Goddard Space Flight Center; the on-line Australia Telescope National Facility (ATNF) pulsar archive; as well as the *ROSAT* all-sky survey from the Max-Planck Institute were all invaluable to this study. The Relativistic Astrophysics Group at *IAR* (DFT, JAC & GER) is supported by *CONICET*, *ANPCT*, and Fundación Antorchas. YMB is supported by the High Energy Astrophysics Division and NASA contract NAS8-39073.

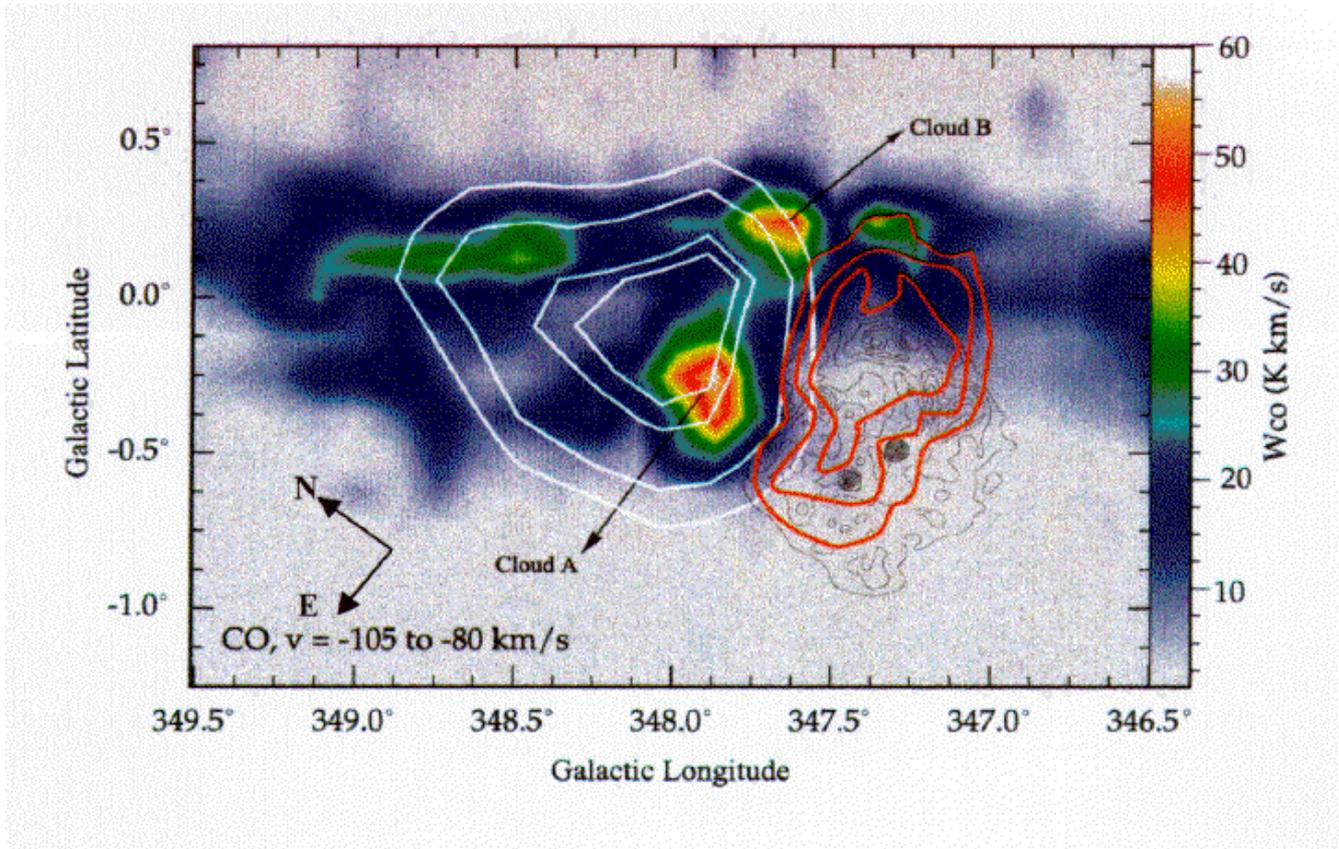

Fig 1: An overlay map in Galactic coordinates showing SNR RX J1713.7-3946 (G347.3-0.5) in grey (*ROSAT* PSPC X-ray) contours from Slane et al. (1999). Red depicts the TeV significance contours from Muraishi et al. (2000). In white are the location probability contours (successively, 50%, 68%, 95% and 99%) of the GeV *EGRET* source 3EG J1714-3857 from Hartman et al. (1999). The color-scale indicates the intensity of CO($J=1\rightarrow0$) emission, and consequently the column density of the ambient molecular cloud, in the LSR velocity interval $v_{lsr}$= -105 to -80 km/sec associated with the SNR, corresponding to a kinematic distance of 6.3±0.4 kpc. The elongated CO emission feature near $(l,b)\sim(348.5,+0.2)$ derives from the large velocity wings of a much more distant (~11.3 kpc) and unrelated cloud centered at $V_{lsr}$= -68 km/sec.

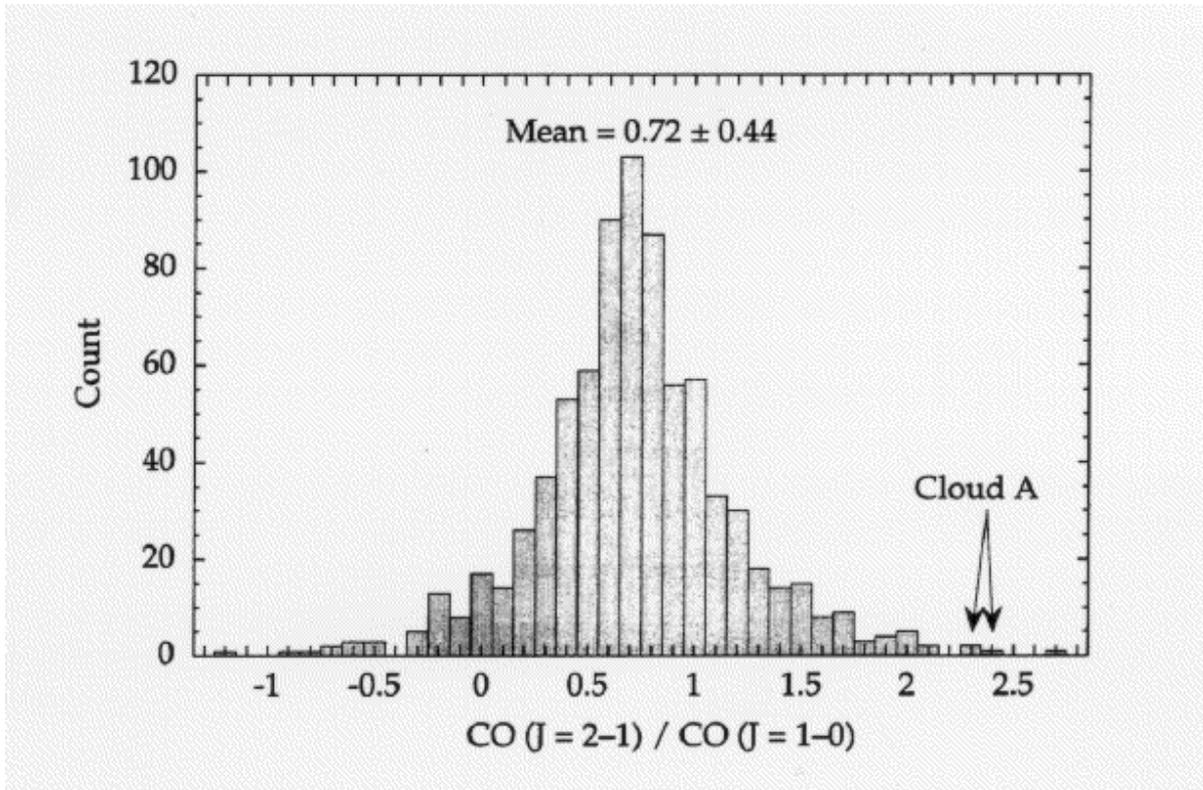

Fig 2: The distribution of all 781 line intensity ratios, $R=\{CO(J=2\rightarrow1)/CO(J=1\rightarrow0)\}$, measured every 15′ in the region from $l=346.5\rightarrow348.5$; $b=-0.5\rightarrow+0.5$, and averaged over 5km/sec bins of velocity between $v_{lsr}=-150$ km/sec $\rightarrow +50$ km/sec. Pixels in which the $CO(J=1\rightarrow0)$ intensity is less than 2.5 times the instrumental noise are excluded. The bins labeled "Cloud A" contain 3 pixels, 2 of which are consistent with the position and velocity of Cloud A: $(l,b,v)=(348.0,-0.25,-85$ km/s$)$, and $(l,b,v)=(348.25,-0.5,-90$ km/s$)$. All other pixels with high $R$ values ($R>1.8$) lie well outside the 95% confidence location contour of 3EG J1714-3857. The mean of the distribution, ~0.72, agrees with the average unexcited value in the Galactic plane (Sakamoto et al., 1995). The dispersion about the mean of ~0.44, results both from the intrinsic scatter in $R$, as well as from instrumental noise, mainly in the $CO(J=2\rightarrow1)$ data. That the latter source dominates is evidenced by some unphysical negative R values which are caused by background subtraction in pixels with very low $CO(J=2\rightarrow1)$ intensity.